\documentclass[amssymb,aps,amsmath,floatfix,prfluids,onecolumn,superscriptaddress,longbibliography]{revtex4-2}
\usepackage[latin1]{inputenc}
\usepackage{mathptmx}
\usepackage[T1]{fontenc}
\pdfoutput=1

\usepackage{graphicx}
\usepackage{amsmath,amssymb,amsfonts, MnSymbol}
\usepackage{mathrsfs}
\usepackage{bm}
\usepackage{hyperref}
\usepackage{bbold}
\usepackage{xcolor}
\usepackage{tikz}
\usepackage{overpic}
\newcommand{\algn}[1]{\begin{align} #1 \end{align}}

\newcommand{\pmat}[1]{\begin{pmatrix} #1 \end{pmatrix}}

\newcommand{\tr}{\text{Tr}}

\newcommand{\taup}{\ensuremath{\tau_\text{p}}}
\newcommand{\tauk}{\ensuremath{\tau_\text{K}}}
\newcommand{\tauc}{\ensuremath{\tau_\text{c}}}

\newcommand{\ve}[1]{\bm{#1}}

\newcommand{\ee}{\ensuremath{\text{e}}}
\newcommand{\ed}{\ensuremath{\text{d}}}
\newcommand{\dd}[1]{\ensuremath{\tfrac{\text{d}}{\text{d} #1}}}

\newcommand{\mc}[1]{\ensuremath{\mathcal{#1}}}
\newcommand{\mbb}[1]{\ensuremath{\mathbb{#1}}}
\newcommand{\ms}[1]{\ensuremath{\mathscr{#1}}}
\newcommand{\st}{\ensuremath{\text{St}}}
\newcommand{\ku}{\ensuremath{\text{Ku}}}
\newcommand{\eqnlab}[1]{\label{eq:#1}}
\newcommand{\seclab}[1]{\label{sec:#1}}
\newcommand{\figlab}[1]{\label{fig:#1}}
\newcommand{\eqnref}[1]{\eqref{eq:#1}}
\newcommand{\Eqnref}[1]{Eq.~\eqref{eq:#1}}
\newcommand{\Eqsref}[1]{Eqs.~\eqref{eq:#1}}

\newcommand{\Secref}[1]{Sec.~\ref{sec:#1}}
\newcommand{\figref}[1]{\ref{fig:#1}}
\newcommand{\Figref}[1]{Fig.~\ref{fig:#1}}
\newcommand{\Figsref}[1]{Figs.~\ref{fig:#1}}

\begin{document}
\title{Caustics in turbulent aerosols form along the Vieillefosse line at weak particle inertia}

\author{Jan Meibohm}
\affiliation{Department of Mathematics, King's College London, WC2R 2LS, United Kingdom}
\author{Kristian Gustavsson}
\affiliation{Department of Physics, Gothenburg University, SE-41296 Gothenburg, Sweden}
\author{Bernhard Mehlig}
\affiliation{Department of Physics, Gothenburg University, SE-41296 Gothenburg, Sweden}

\begin{abstract}
Caustic singularities of the spatial distribution of particles in turbulent aerosols enhance collision rates and accelerate coagulation.  Here we investigate how and where caustics form at weak particle inertia, by analysing a three-dimensional Gaussian statistical model for turbulent aerosols in the persistent limit, where the flow varies slowly compared with the particle relaxation time. In this case, correlations between particle- and fluid-velocity gradients are strong, and caustics are induced by large, strain-dominated excursions of the fluid-velocity gradients. These excursions must cross a characteristic threshold in the plane spanned by the invariants $Q$ and $R$  of the fluid-velocity gradients. Our method predicts that the most likely way to reach this threshold is by a unique ``optimal fluctuation''   that propagates along the Vieillefosse line, $27R^2/4 +Q^3=0$. We determine the shape of the optimal fluctuation as a function of time and show that it is dominant in numerical statistical-model simulations even for moderate particle inertia.
\end{abstract}
\maketitle
\section{Introduction}
In turbulent aerosols, particle inertia allows heavy particles to detach from the flow and generate folds over configuration space, so-called caustics~\cite{Cri92,Fal02,Wil05,Mei20a}. Near such caustic folds, phase-space neighbourhoods partially focus onto configuration space, leading to transient divergencies of the spatial particle-number density, analogous to caustics in geometrical optics \cite{Ber80}.
At the same time, caustics delineate regions in configuration space where the particle velocities are multi valued, a phenomenon known as the sling effect \cite{Fal02,Fal07c}. It leads to anomalously large relative particle velocities~\cite{Fal02,Bec10,Sal12,Gus14c}, increased collision rates \cite{Vol80,Sun97,Fal02,Vos14,Pum16}, and large collision velocities \cite{Bec05,Bec10,Sal12,Gus14c}. The latter can significantly affect collision outcomes \cite{Wil08,Win12}.

The process of caustic formation depends sensitively on the characteristic, dynamical time scales of the fluid and of the particles~\cite{Gus16}. For a dilute, monodisperse suspension of dense, identical particles, there are three time scales: the Eulerian correlation time $\tauc$ of the flow, its Lagrangian correlation time $\tauk$, measured along the paths of fluid elements, and the particle relaxation time $\taup$, the time it takes the inertial-particle velocity to relax back to the fluid velocity. The particle dynamics is characterised by two dimensionless numbers: The Stokes number $\st\equiv\taup/\tauk$ is a dimensionless measure of particle inertia, while the Kubo number $\ku\equiv\tauc/\tauk$ determines the degree of persistence of the flow.

The rate of caustic formation has been measured using direct numerical simulation of turbulence \cite{Fal07c,Bha22} and kinematic turbulence models \cite{Duc09,Men11a}.  Yet the mechanisms of caustic formation are understood only in certain idealised models and limiting cases. In the so-called white-noise limit~\cite{Gus16}, the fluid velocities seen by the particles are assumed to fluctuate rapidly ($\ku\ll1$), and particle inertia is assumed to be large ($\st\gg1$). For Gaussian random velocity fields in this limit, caustic formation can be described within a diffusion approximation which predicts that the particle dynamics depends sensitively on the (small) value of the Kubo number~\cite{Gus16}.
 
Aerosol models with Gaussian random velocity fields that realistically mimic homogeneous and isotropic turbulence, by contrast, require Kubo numbers of order $10$, and are thus in a parameter regime where the particle dynamics depends only weakly upon $\ku$~\cite{Gus16}. This is a motivation for analysing the opposite, so-called persistent limit of large Kubo numbers and small Stokes numbers. It corresponds to slowly varying fluid velocities and short particle relaxation times.
Model calculations in one~\cite{Der07,Mei17,Mei19} and two~\cite{Mei21a} spatial dimensions show that caustics form in the persistent limit when the matrix of fluid-velocity gradients reaches a large threshold. Consequently, the most likely way for caustics to form in this limit is by rare, large fluctuations of the fluid-velocity gradients that exceed this threshold. For small $\st$, these fluctuations are dominated by a single optimal fluctuation that can be calculated using methods from large deviation theory~\cite{Var84,Fre84,Tou09}, recently reviewed in, e.g.,~\cite{Gra15,Ass17}.

In this paper, we compute the optimal fluctuation of the fluid-velocity gradient matrix $\mbb A$, and the corresponding most likely (optimal) path to caustic formation for an incompressible turbulent flow, by analysing a three-dimensional statistical model in the persistent limit. We first characterise $\mbb{A}$ in terms of its invariants~$Q=-\tr(\mbb{A}^2)/2$ and $R=-\tr(\mbb{A}^3)/3$~\cite{Cho90,Sor94} and show that it must cross a characteristic threshold line in the $Q$-$R$ plane for a caustic to form. We then employ large-deviation techniques~\cite{Gra15,Ass17} to determine how this threshold is reached. Our method predicts that the optimal fluctuation, given by the most likely way to reach the threshold, moves along the positive branch ($R>0$) of the so-called Vieillefosse line~\cite{Vie82,Vie84,Can92,Men11b},
\algn{\eqnlab{VFL}
\frac{27}{4}R^2 +Q^3=0\,,
}
which forms the boundary between extensional and rotational flow configurations. The optimal fluctuation is vorticity-free and unique up to similarity transformations, but exact only for $\st\ll1$. Nevertheless, numerical simulations of a Gaussian statistical model show that it remains dominant even at $\st$ numbers as large as $\st\approx 0.3$, suggesting a distinct way of caustic formation in three spatial dimensions that can be tested in experiments~\cite{Bew13}. In high-Reynolds-number turbulence, the probability distribution of Lagrangian fluid-velocity gradients is strongly skewed along the positive branch of the Vieillefosse line~\cite{Can93,Che99,Che06}. This means that caustic formation is expected to be significantly higher in actual turbulent flows compared with models that neglect this skewness.
\section{Problem formulation}
The dynamics of a spherical particle in a fluid-velocity field $\ve u(\ve x,t)$ is well approximated by Stokes's law \cite{Bat00}
\algn{\eqnlab{stokes}
	\dd{t}\ve x(t) = \ve v(t)\,, \qquad  \dd{t}\ve v(t) = \taup^{-1}\{\ve u[\ve x(t),t]-\ve v(t)\}\,,
}
given that the particle is small enough and much denser than the fluid. Here, $\ve x$ and $\ve v$ denote particle position and velocity. The particle relaxation time $\taup=2a^2 \rho_\text{p}/(9\rho_\text{f}\nu)$ is a function of the particle size $a$, the kinematic viscosity $\nu$ of the fluid, and of the particle and fluid densities, $\rho_\text{p}$ and $\rho_\text{f}$, respectively.

In $d$ spatial dimensions, a caustic occurs when the separation vectors between $d+1$ nearby particles partially align, so that the spatial volume $\mathscr{\hat V}(t)=|\det\mbb{J}(t)|$, occupied by the particles, momentarily collapses to zero. Here, $\mbb{J}$ denotes the spatial deformation tensor $J_{ij}[\ve x(t_0),t] = \partial x_{i}(t)/\partial x_{j}(t_0)$. Caustic formation is closely related to the dynamics of the particle-velocity gradients~\cite{Fal02,Gus16}, written in dimensionless form as
\algn{\eqnlab{model}
	\st\dd{t}{\mbb Z}(t)=-\mbb Z(t)-\mbb Z(t)^2+ \mbb A(t)\,,
}
with $\mbb Z(t_0) = \mbb A(t_0)$ initially. Here, $Z_{ij}(t) = \taup\partial v_i(t)/\partial x_j(t)$  and $A_{ij}(t) = \taup\partial u_i(t)/\partial x_j(t)$, are dimensionless matrices of Lagrangian particle-velocity gradients and fluid-velocity gradients, respectively, evaluated at the particle position, i.e., $\mbb Z(t) = \mbb Z[\ve x(t),t]$ and $\mbb A(t) = \mbb A[\ve x(t),t]$. Time $t$ in \Eqnref{model} is measured in units of the Kolmogorov time, $\tauk = \taup\langle \tr\mbb{A}\mbb{A}^{\sf T}\rangle_f^{-1/2}$, where $\langle\ldots\rangle_f$ denotes a Lagrangian average along the paths of fluid elements. The particle-velocity gradients $\mbb{Z}$ are related to the spatial volume $\ms{\hat V}(t)$ through
\algn{\eqnlab{vzeqn}
	\mathscr{\hat V}(t) = \mathscr{\hat V}(t_0)\exp\int^t_{t_0}\!\!{\rm d}s\,\tr\mbb{Z}(s)\,.
}
Consequently, a necessary condition for caustic formation, $\mathscr{\hat V}(t)\to0$, is that $\tr\mbb{Z}(t)$ escapes to negative infinity.

In the small-$\st$ limit, the particle dynamics, \Eqsref{stokes} and \eqnref{model}, occurs on much shorter time scales than the dynamics of $\mbb{A}$, so the problem can be treated using the persistent limit~\cite{Der07,Mei19,Mei21a}. In this limit, changes of \mbb{A} in \Eqnref{stokes} are adiabatic, i.e., \mbb{A} remains effectively constant~\cite{Mei19,Mei21a} as $\mbb{Z}$ varies. Whenever \Eqnref{model} has a globally attracting stable fixed point $\mbb{Z}_*$, where
\algn{\eqnlab{FPE}
	\mathbb{F}(\mbb{Z}_*,\mbb{A})\equiv -\mbb Z_*-\mbb{Z}_*^2+ \mbb  A=0\,,
}
$\mbb{Z}$ rapidly relaxes to  $\mbb{Z}_*$ within a short transient time of order $\st$. In this case, $\tr\mbb{Z}$ remains finite and no caustics form. In the absence of stable fixed points, by contrast, the dynamics drives the particle gradient matrix $\mbb{Z}$ to infinity, so $\tr\mbb{Z}\to-\infty$, implying the formation of a caustic.
 
In summary, at small $\st$ caustics form through bifurcations of the fixed-point equation \eqnref{FPE}, induced by large but adiabatically slow excursions of $\mbb{A}$. The probability and shape of these excursions are determined by optimal fluctuation theory~\cite{Gra15,Ass17}, which predicts the most likely excursion that leads to a bifurcation and thus to a caustic.
\section{Fixed-point analysis}
To find the bifurcations of the fixed points $\mbb{Z}_*(\mbb{A})$ as $\mbb{A}$ changes adiabatically, we note that \Eqnref{FPE} satisfies
\algn{\eqnlab{equivariance}
	\mbb{F}(\mbb{P}^{-1}\mbb{Z}\mbb{P},\mbb{P}^{-1}\mbb{A}\mbb{P}) = \mbb{P}^{-1}\mbb{F}(\mbb{Z},\mbb{A})\mbb{P}\,,
}
where $\mbb{P}$ is an arbitrary invertible matrix. As a consequence, if $\mbb{Z}_*(\mbb{A})$ is a fixed point, then so is $\mbb{P}^{-1}\mbb{Z}_*(\mbb{P}^{-1}\mbb{A}\mbb{P})\mbb{P}$. Due to the symmetry \eqnref{equivariance}, the set of fixed points and their stability are invariant under similarity transformations~\cite{Gol88}, and thus only depend on the invariants of $\mbb{A}$. For incompressible flow, $\mbb{A}$ is traceless and obeys the characteristic equation ~\cite{Cay58,Ham53} 
\algn{\eqnlab{CE}
	\mbb{A}^3  - \tfrac12 \tr(\mbb{A}^2)\mbb{A} - \det(\mbb{A})\mbb{1}=0
}
with invariants~\cite{Cho90}
\algn{\eqnlab{ainv}
	Q(\mbb{A}) = - \frac12\tr(\mbb{A}^2)\,,\qquad R(\mbb{A}) = -\det(\mbb{A})=-\frac13\tr(\mbb{A}^3)\,.
}
An analogous characteristic equation holds for $\mbb{Z}_*$, but with a non-vanishing trace term. We characterise the fixed points $\mbb Z_*$ by evaluating $\cal Z_* \equiv \tr\mbb Z_*$. To this end, we raise the fixed-point equation $\mbb{A} = \mbb{Z}_* + \mbb{Z}_*^2$  to the second and third powers and take the trace. This gives $Q$ and $R$ in terms of $\tr(\mbb{Z}_*^n	)$, with $n\leq6$. From the characteristic equation of $\mbb{Z}_*$ we obtain, by multiplication with $\mbb{Z}_*$ and taking the trace, expressions for $\tr(\mbb{Z}_*^n)$ that we substitute into the equations for $Q$ and $R$. In this way, we end up with one single equation for $\mc{Z}_*$:
\algn{\eqnlab{qreqn}
	-16 [2 \mc{Z}_*+3]^2 R =&	\left\{\mc{Z}_* [\mc{Z}_*+1]^2 [\mc{Z}_*+2]-4 Q\right\} \left\{[\mc{Z}_*+1] [\mc{Z}_*+2]^2 [\mc{Z}_*+3]-4 Q\right\}\,.
}
From \Eqnref{qreqn} we conclude that possible fixed-point values $\mc{Z}_*$ depend only on the invariants $Q$, and $R$ of $\mbb{A}$. Equation~\eqnref{model} is unstable in regions in the $Q$-$R$ plane where \Eqnref{qreqn} has no solution, allowing $\tr\mbb{Z}$ to escape to $-\infty$, so that a caustic forms. 

In general, regions with different numbers of solutions of \Eqnref{qreqn} (ranging from zero to eight, see discussion below) are separated by saddle-node bifurcations of $\mbb{F}(\mbb{Z}_*,\mbb{A})=0$, where fixed points are created or destroyed~\cite{Str18}. Exceptions are bifurcations at isolated points or on symmetry lines where other, e.g. higher-order, bifurcations may occur~\cite{Str18}. To find the locations of the saddle-node bifurcations in the $Q$-$R$ plane, we introduce the coordinate $z_*=(\mathcal{Z}_*+3/2)^2$, because it turns \Eqnref{qreqn} into a quartic polynomial equation in $z_*$. Saddle-node bifurcations occur at the simultaneous positive roots of this polynomial and of its derivative with respect to $z_*$. Solving for these roots, we find two bifurcation lines, shown as solid lines in \Figref{phasediag}. The first one (orange line in \Figref{phasediag}) is the Vieillefosse line \eqnref{VFL} for $R\leq1/256$ which distinguishes rotational and extensional regions in the flow. 
We note that its positive branch ($R>0$) plays an important role for the Lagrangian dynamics of fluid-velocity gradients. Vieillefosse~\cite{Vie82,Vie84} showed that Lagrangian fluid-velocity gradients described by the inviscid Navier-Stokes equations self-amplify along the positive branch of \Eqnref{VFL}. When the anisotropic portion of the pressure Hessian is neglected, this self-amplification leads to a finite-time divergence of Lagrangian fluid-velocity gradients along this branch~\cite{Vie82,Vie84,Can92}. Although the divergence does not occur in more realistic approximations~\cite{Gir90,Che99,Che06}, the self-amplification effect explains  why the distribution of Lagrangian fluid-velocity gradients in Navier-Stokes turbulence is strongly skewed towards the positive branch of \Eqnref{VFL}~\cite{Men11b}.

The second bifurcation curve (red line in \Figref{phasediag}) is given by
\algn{\eqnlab{bifline}
	Q = 4R-1/16\,\quad\text{for}\quad R\geq-1/32\,.
}
The two bifurcation lines meet at $R=-1/32$ (not shown) and at $R=1/256$ (red square in \Figref{phasediag}). They divide the $Q$-$R$ plane into regions where \Eqnref{qreqn} has different numbers of solutions.
\begin{figure}
	\centering
	\includegraphics[]{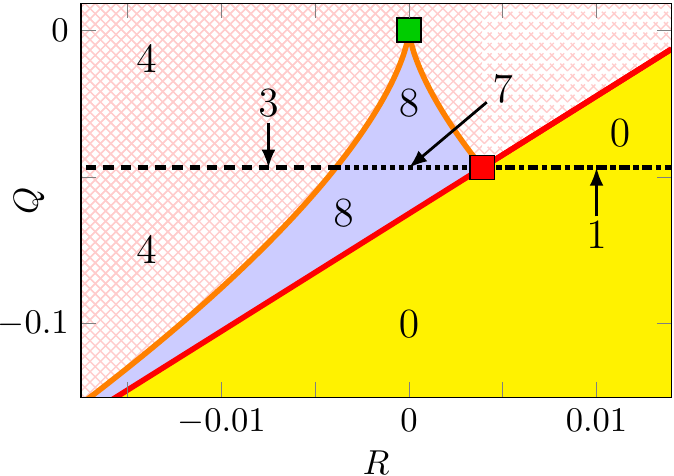}
	\caption{Phase diagram with regions of different numbers of solutions of \Eqnref{qreqn} as a function of $Q$ and $R$. The lines show the bifurcations described in the text. The green square $(R,Q)=(0,0)$ shows the most likely gradient configuration at steady state, while for infinitesimal particle inertia the threshold \eqnref{bifline} is reached at the red square $(R,Q)=(1/256,-3/64)$, see \Secref{optfluct}.}\figlab{phasediag}
\end{figure}
In addition, transcritical bifurcations, which conserve the number of fixed points~\cite{Str18}, occur on the symmetry line (broken lines in \Figref{phasediag}):
\algn{
Q=-3/64\,,
}
where the $z_*$ derivative of \Eqnref{qreqn} vanishes. Importantly, the bifurcation line \eqnref{bifline} separates the region without any fixed points (yellow region) from the rest of the $Q$-$R$ plane (blue and crosshatched regions).  The latter regions contain a finite number of fixed points and we find numerically that one of them is always stable and smoothly connected to $\mbb{Z}_*\approx \mbb{A}$ close to the origin.
 
In summary, at small $\st$ we can invoke time-scale separation. While $\mbb Z$ relaxes rapidly, $\mbb A$ changes slowly. As soon as $\mbb A$ crosses the threshold line \eqnref{bifline} and enters the yellow region in \Figref{phasediag}, the $\mbb Z$ dynamics becomes unstable. As a result, a caustic forms since $\tr\mbb{Z}(t)\to-\infty$ in a short time.
\newpage
\section{Optimal fluctuation}\seclab{optfluct}
At small particle inertia caustics are rare, because typical fluctuations of $\mbb{A}$ along particle paths are of order $\st\ll1$, much smaller than the threshold \eqnref{bifline} which is of order unity. While there are typically many ways of similar probability to realise a typical fluctuation (here of order $\st\ll1$), this is not the case for rare, large fluctuations. Instead, it is a general theme in large-deviation theory~\cite{Var84,Fre84,Tou09} that rare events (here: fluctuations of order unity) are often dominated by unique, optimal fluctuations, that correspond to the most likely realisations of these events. Here we use large-deviation techniques developed to characterise such rare events~\cite{Gra15,Ass17} to determine the optimal fluctuation for $\mbb{A}$ to reach the threshold, starting from the most likely initial value $Q=R=0$. 
\subsection{Matrix basis}\seclab{matbas}
In homogeneous and isotropic turbulence, the elements $A_{ij}$ of $\mbb{A}$ are correlated~\cite{Fal01} which makes calculations cumbersome. It is therefore convenient to represent $\mbb{A}$ in a basis of matrices $\mbb{e}_1,\ldots,\mbb{e}_8$, $\mbb A(t) = \sum_{i=1}^8A_i(t) \mbb{e}_i$, chosen in such a way that the random processes $A_i(t)$ are uncorrelated. The basis we use in the following is orthonormal with respect to the inner product $\langle \mbb M,\mbb N\rangle \equiv \tr(\mbb{MN})/2$ such that $\langle \mbb{e}_i,\mbb{e}_j\rangle = g_{ij}$, where
\algn{
	\mbb{g} =\pmat{-1&&&&&&&\\&-1&&&&&&\\&&-1&&&&&\\&&&\phantom{-}1&&&&\\&&&&\phantom{-}1&&&\\&&&&&\phantom{-}1&&\\&&&&&&\phantom{-}1&\\&&&&&&&\phantom{-}1}.
}
The basis elements $\mbb{e}_1$, $\mbb{e}_2$, and $\mbb{e}_3$ span the space of traceless antisymmetric matrices and can be written
\algn{\eqnlab{basis}
        \mbb{e}_1 =&
\pmat{
 0 & 0 & 0 \\
 0 & 0 & -1 \\
 0 & 1 & 0}\,,
 \quad
        \mbb{e}_2 =
\pmat{
 0 & -1 & 0 \\
 1 & 0 & 0 \\
 0 & 0 & 0}\,,
 \quad
        \mbb{e}_3 =
\pmat{
 0 & 0 & -1 \\
 0 & 0 & 0 \\
 1 & 0 & 0}\,.
}
The remaining $\mbb{e}_i, i=4,\ldots,8$, spanning the space of traceless symmetric matrices, read
\algn{\eqnlab{basis2}
        \mbb{e}_4 =
\pmat{
 0 & 0 & 1 \\
 0 & 0 & 0 \\
 1 & 0 & 0}\,,
 \quad
        \mbb{e}_5 =
\pmat{
 0 & 0 & 0 \\
 0 & 0 & 1 \\
 0 & 1 & 0}\,,
 \quad
        \mbb{e}_6 =
\pmat{
 1 & 0 & 0 \\
 0 & -1 & 0 \\
 0 & 0 & 0}\,,
 \quad
        \mbb{e}_7 =
\pmat{
 0 & 1 & 0 \\
 1 & 0 & 0 \\
 0 & 0 & 0}\,,
 \quad
        \mbb{e}_8 =\frac1{\sqrt{3}}
\pmat{
 1 & 0 & 0 \\
 0 & 1 & 0 \\
 0 & 0 & -2}\,.
 &
}
Consequently, the antisymmetric (vorticity) and symmetric (strain) parts of $\mbb{A}$ are expanded as
\algn{\eqnlab{trcorr}
	\mbb{O}(t) =  \sum_{i=1}^3O_i(t)\mbb{e}_i\,,\qquad \text{and}\qquad \mbb{S}(t) = \sum_{i=1}^5S_i(t)\mbb{e}_{i+3}\,,
}
respectively, where $O_i = A_i$ and $S_i = A_{i+3}$. Note that since $\mbb{e}_i^{\sf T} = \sum_{j=1}^{8}g_{ij}\mbb{e}_j$, one has $\langle \mbb{e}_i,\mbb{e}_i^{\sf T}\rangle = \delta_{ij}$. In addition to $\mbb{e}_1,\ldots,\mbb{e}_8$ a complete matrix basis also includes the unit matrix $\mbb e_0=\mbb 1$, which is, however, not needed in the following, because $\mbb A$ is traceless.
\subsection{Correlation functions of fluid-velocity gradients}
For homogeneous and isotropic turbulence, the entries of $\mbb{O}$ and $\mbb{S}$ have zero mean and correlation functions~\cite{Bru97,Zai03}
\algn{\eqnlab{tenscorr}
	\langle \mbb{O}_{ik}(t)\mbb{O}_{jl}(t')\rangle_p = \sigma_O^2 C^O_{ijkl}f_O(t-t')\,,\qquad \langle \mbb{S}_{ik}(t)\mbb{S}_{jl}(t')\rangle_p = \sigma_S^2 C^S_{ijkl}f_S(t-t')\,,\qquad \langle \mbb{O}_{ik}(t)\mbb{S}_{jl}(t')\rangle_p = 0\,,
}
where the angular brackets $\langle\cdot\rangle_p$ denote a steady-state average along inertial-particle trajectories. The correlation functions $f_O$ and $f_S$ are normalised such that $f_O(0)=f_S(0)=1$, and are well approximated by exponential functions with different correlation times of order $\tauk$~\cite{Gir90,Vin07}. The structure of the rotationally covariant tensors $C^S$ and $C^O$ in \Eqsref{tenscorr} follows from isotropy and incompressibility~\cite{Gir90,Bru97}:
\algn{
	C^O_{ijkl}=\delta_{ij}\delta_{kl} - \delta_{il}\delta_{jk}\,,\qquad C^S_{ijkl}=\delta_{ij}\delta_{kl} + \delta_{il}\delta_{jk} - \frac23\delta_{ik}\delta_{jl}\,.
}
Along the paths of fluid elements, the Lagrangian flow is homogeneous, which implies that~\cite{Gir90,Bru97}
\algn{
\langle\tr[\mbb{O}^{\!\sf T}\!\!(t)\mbb{O}(t)]\rangle_f = \langle\tr[\mbb{S}^{\!\sf T}\!\!(t)\mbb{S}(t)]\rangle_f = \frac12\langle\tr[\mbb{A}^{\!\sf T}\!\!(t)\mbb{A}(t)]\rangle_f = \frac{\st^2}2\,.
}
Along inertial particle trajectories, however, preferential concentration~\cite{Squ91,Eat94} breaks homogeneity~\cite{Zai09}, so that $\langle\tr[\mbb{O}^{\!\sf T}\!\!(t)\mbb{O}(t)]\rangle_p < \langle\tr[\mbb{S}^{\!\sf T}\!\!(t)\mbb{S}(t)]\rangle_p$~\cite{Max87}. We therefore write the variances $\sigma_O$ and $\sigma_S$ in \Eqnref{trcorr} as 
\algn{\eqnlab{vars}
	\sigma_O^2 = \frac16\st^2 C_O(\st)\,,\qquad \text{and}\qquad \sigma_S^2 = \frac1{10}\st^2 C_S(\st)\,,
}
where $C_O(\st)\leq1/2$ and $C_S(\st)\geq1/2$ are $\st$-dependent functions with $\lim_{\st\to0}C_O(\st)=\lim_{\st\to0}C_S(\st)=1/2$ in the inertialess limit. The prefactors in \Eqnref{vars} account for the three and five independent components of $\mbb{O}$ and $\mbb{S}$, respectively. In terms of the orthonormal basis introduced in \Secref{matbas}, the correlation functions \eqnref{tenscorr} are conveniently expressed as
\algn{\eqnlab{oscorr}
	\langle O_i(t) O_j(t') \rangle_p = \delta_{ij}\sigma^2_Of_\text{O}(t-t')\,,\qquad \langle S_i(t) S_j(t') \rangle_p = \delta_{ij}\sigma^2_Sf_\text{S}(t-t')\,,\qquad \langle O_i(t) S_j(t') \rangle_p = 0\,.
}
Equation~\eqnref{oscorr} makes explicit that the elements $A_i(t)$ are mutually uncorrelated processes with variances $\sigma_O$ and $\sigma_S$.
\subsection{Gaussian approximation}
The complete statistical information about $\mbb{A}(t)$ is contained in the cumulant-generating functional~\cite{Mon71} of the process
\algn{\eqnlab{cgfmat}
	\Lambda(\mbb{G},t)=&\ln\left\langle\exp\left\{\int_{t_0}^t\!\!\ed s\langle\mbb{A}(s),\mbb{G}(s)\rangle\right\}\right\rangle_p\,.
}
Here $\mbb{G}$ is a traceless matrix-valued test function. An expansion in powers of $\mbb{G}(s)=\sum_{i=1}^8 G_i(s) \mbb{e}_i$ generates the cumulants of $\mbb{A}(t)$, which we write in terms of $A_i(t)$. Since the correlation functions for the processes $A_i(t)$ decouple in the orthonormal matrix basis $\mbb{e}_i$ [see \Eqnref{oscorr}] we find to second order in $\mbb{G}$
\algn{\eqnlab{cgfvec}
		\Lambda(\mbb{G},t)=\sum_{i=1}^8\int_{t_0}^t\int_{t_0}^t\!\!\ed s\,\ed s' \sigma_i^2 f_i(s-s')G_i(s) G_i(s') + \mc{O}(\mbb{G}^3)\,,
}
with $\sigma^2_i f_i(t-t')=\sigma^2_Of_O(t-t')$ for $i=1,\ldots,3$ and $\sigma^2_if_i(t-t')=\sigma^2_Sf_S(t-t')$ for $i=4,\ldots,8$. The orders $\mc{O}(\mbb{G}^3)$ in \Eqnref{cgfvec} contain terms of the form $\langle A_i(s)A_j(t)A_k(u)\rangle_p$, $\langle A_i(s) A_j(t) A_k(u) A_l(v)\rangle_p$, and so on. In a mean-field (or Gaussian) approximation, we decompose the higher order correlation functions in terms of the second  and first-order ones. This gives $\langle A_i(s)A_j(t)A_k(u)\rangle_p\approx\langle A_i(s)A_j(t)\rangle_p \langle A_k(u)\rangle_p+\ldots = 0$ and
\algn{
	\langle A_i(s) A_j(t) A_k(u) A_l(v)\rangle_p \approx \langle A_i(s) A_j(t)\rangle_p\langle A_k(u) A_l(v)\rangle_p + \ldots\,,
}
where the ellipsis denotes all permutations of the arguments. Applied to all orders, this approximation makes the terms $\mc{O}(\mbb{G}^3)$ in \Eqnref{cgfvec} vanish so that the individual processes $A_i(t)$ become Gaussian. Although the fluid-velocity gradients in the dissipation range of isotropic and homogeneous turbulence are well known not to be Gaussian~\cite{Mon75, Fri97,Pop00}, Gaussian models for turbulence have proven to provide valuable information about the behaviour of turbulent aerosols~\cite{Zai09,Gus16,Pum16}. Uncorrelated Gaussian processes are independent, as can be seen from the fact that $\Lambda(\mbb{G},t)$ decomposes completely when the $\mc{O}(\mbb{G}^3)$ term in \Eqnref{cgfvec} is absent. The mutual independence of the processes $A_i(t)$ is a crucial ingredient for the calculations that follow.
\subsection{Threshold probability}
For the stationary Gaussian process $A_i(t)$, the probability for reaching a given value $A_i$ at a given time $t$, reads
\algn{\eqnlab{transAi}
	P[A_i(t)=A_i] \propto\ee^{-S(A_i)/\sigma_i^2}\,,
}
where we omitted the normalisation prefactor. Here, $S(A_i) = A_i^2/2$ denotes the quadratic action associated with reaching $A_i$. Due to the mutual independence of the $A_i(t)$, we obtain the probability for $\mbb{A}$ to reach $\mbb{A}\equiv\sum_{i=1}^8A_i \mbb{e}_i$ as the product of \Eqnref{transAi} over $i$,
\algn{\eqnlab{transAmat}
	P[\mbb{A}(t)=\mbb{A}] = \prod_{i=1}^8 P[A_i(t)=A_i] \propto \ee^{-S(\mbb{A})/\sigma_S^2}\,,
}
characterised by the weighed sum $S(\mbb{A})$ of actions $S(A_i)$ over $i$:
\algn{\eqnlab{action}
	S(\mbb{A}) = \frac12\left(\frac{\sigma_S}{\sigma_O}\right)^2\sum_{i=1}^3 A_i^2 + \frac12\sum_{i=4}^8 A_i^2= -\frac12\left(\frac{\sigma_S}{\sigma_O}\right)^2\!\left\langle\mbb{O},\!\mbb{O}\right\rangle+ \frac12\left\langle\mbb{S},\mbb{S}\right\rangle\,.
}
Since $\sigma_S, \sigma_O \sim \st$, \Eqnref{transAmat} and \eqnref{action} imply that the exponent in \Eqnref{transAmat} is proportional to $\st^{-2}$ for small values of $\st$. Also note that $S(\mbb{A})$ is quadratic in $A_i$. Both observations are a consequence of approximating  $A_i(t)$ as Gaussian processes.

\subsection{Optimal threshold configuration}
To obtain the most likely way for the fluid-velocity gradients to reach the threshold \eqnref{bifline} and induce a caustic, we must minimise the action \eqnref{action} under the constraint \eqnref{bifline}. To this end, we multiply the constraint with a Lagrange multiplier $\lambda$, and
add the product to $S(\mbb{A})$ to form the Lagrange function
\algn{\eqnlab{lagr}
	\ms{L}(\mbb{A})=& S(\mbb{A}) - \lambda \left[	Q(\mbb{A})-4R(\mbb{A})+\frac1{16}	\right]\,.
}
Minimising $\ms{L}$ over $\mbb{A}$ gives the minimiser $\mbb{A}^*$, the most likely configuration of fluid-velocity gradients at the threshold \eqnref{bifline}. 
The symmetry of $\mbb F$ under similarity transformations [see \Eqnref{equivariance}] constrains this minimiser. In order to deduce how,
we consider infinitesimal similarity transformations $\mbb{A}\to\mbb{A'}=\mbb{P}^{-1}\mbb{A}\mbb{P}$ with $\det\mbb{P}=1$, generated by traceless matrices
\algn{\eqnlab{matAtrafo}
        \mbb{A'}\approx\mbb{A} + \delta \mbb{A}\,,\qquad\delta\mbb{A}=\sum_{i=1}^8[\mbb{A},\mbb{e}_i] \theta_i \,,
}
where $[\mbb{M},\mbb{N}]=\mbb{MN}-\mbb{NM}$ denotes the commutator between two matrices, and the factors $|\theta_i|\ll1$ parametrise the infinitesimal transformation. For the symmetric and antisymmetric parts of $\mbb{A}'$, the transformation \eqnref{matAtrafo} implies $\mbb{S}' = \mbb{S} + \delta\mbb{S}$ and $\mbb{O}' = \mbb{O} + \delta \mbb{O}$ with
\algn{
	\delta\mbb{S}=\sum_{i=1}^3[\mbb{S},\mbb{e}_i] \theta_i + \sum_{i=4}^8[\mbb{O},\mbb{e}_i] \theta_i\,,\qquad \delta\mbb{O}=\sum_{i=1}^3[\mbb{O},\mbb{e}_i] \theta_i + \sum_{i=4}^8[\mbb{S},\mbb{e}_i] \theta_i\,.
}
The Lagrange function $\ms{L}(\mathbb{A})$ in Eq. (27) must be invariant under similarity transformations of the minimizer $\mbb{A}^*$, i.e., $\delta \ms{L}(\mbb{A}) = \ms{L}(\mbb{A'})-\ms{L}(\mbb{A})=0$ at $\mbb{A} = \mbb{A}^*$, which requires that $\delta S(\mbb{A}) = S(\mbb{A}') - S(\mbb{A}) = 0$ at the minimiser. Applying the infinitesimal transformation~\eqnref{matAtrafo} to $S(\mbb{A})$, however, we find
\algn{
	\delta S(\mbb{A}) = \left[\left(\frac{\sigma_S}{\sigma_O}\right)^2 + 1\right]\sum_{i=4}^8\left\langle[\mbb{S},\mbb{O}],\mbb{e}_i\right\rangle\theta_i\,.
}
This expression vanishes, i.e., $S(\mbb{A})$ is invariant, only in two cases: First, $\theta_i=0$ for $i=4,\ldots,8$ which corresponds to the case when $\mbb{P}$ is a rotation $\mbb{P}^{\sf T} = \mbb{P}^{-1}$. The second case is $[\mbb{S},\mbb{O}]=0$ which occurs when either $\mbb{O}=0$ or $\mbb{S}=0$. It can be checked that the constraint~\eqnref{bifline} is invariant under all similarity transformations, not only rotations, because $\delta Q(\mbb{A}) = \delta R(\mbb{A}) = 0$ for all $\theta_i$. Hence, the constraint requires that $\delta S(\mbb{A})=0$ also for finite $\theta_i$, $i=4,\ldots,8$, meaning that either $\mbb{O}=0$ or $\mbb{S}=0$. However, $\mbb{S}=0$, and thus $\mbb{A}=\mbb{O}$, leads to $R(\mbb{O})=0$ and $Q(\mbb{O})\geq 0$, so the constraint~\eqnref{bifline} cannot be satisfied for any real $\mbb{O}$. This in turn, shows that the vorticity contribution $\mbb{O}^*$ to the optimal threshold configuration $\mbb{A}^*$ must vanish, $\mbb{O}^*=0$, and that $\mbb{A}^*$ must be a pure strain $\mbb{A}^*=\mbb{S}^*$.

In order to determine the components of $\mbb{S}^*$, we express $\mbb{A}$ in \Eqnref{lagr} in terms of $A_i$, and minimise~\Eqnref{lagr} over $A_4,\ldots,A_8$, while $A^*_1=A_2^*=A^*_3=0$. This leads to two solutions with equal actions $S(\mbb{A}^*) = 3/64$, which fix the minimising configuration $A^*_i$ up to equivalence under similarity transformations. As a consequence of this symmetry, the minimisers are not isolated points but higher-dimensional manifolds. All minimisers found in this way are related by similarity transformations, so that we may choose particular representatives to gain insight into the most likely threshold configuration $\mbb{A}^*$. The simplest representatives correspond to diagonal $\mbb{A}^*$, and they show that $\mbb{A}^*$ takes two positive eigenvalues equal to $1/8$, and one negative eigenvalue equal to $-1/4$ of twice the magnitude. Similarity transformations only affect the order of eigenvalues, so we may put all minimisers into the ordered diagonal form
\algn{
	\mbb{A}^* =\frac14\pmat{1/2&&\\&1/2&\\&&-1}
}
by a suitable transformation. Expressed in the matrix basis, this corresponds to the simple configuration $A^*_8=(1/4)\sqrt{3}/2$, and $A^*_1=A^*_2=\ldots=A^*_7=0$.

Up to this point, our analysis provides the most likely threshold configuration that induces a caustic in the small-$\st$ limit. The strict limit, however, is difficult to approach in numerical simulations, since the threshold probability~\eqnref{transAmat} is exponentially suppressed in $\st$.
\subsection{Small-\texorpdfstring{$\st$}{St} correction}
 We now discuss how to adjust the theory to incorporate the main next-to-leading order effects in $\st\ll1$, in an approach similar to that used in two spatial dimensions in Ref.~\cite{Mei21a}.
Our starting point is \Eqnref{model}. Since caustics form on time scales of order $\st$, we require the threshold \eqnref{bifline} to be exceeded for a finite time, such as to leave the dynamics~\eqnref{model} sufficient time to form a caustic \cite{Mei21a}. To account for this, we introduce a $\st$-dependent threshold $A_\text{th}(\st)=1/4+o(\st^0)$ \cite{Mei21a} that consists of the threshold determined above plus a small positive correction $o(\st^0)$ that vanishes as $\st\to0$. Although this $o(\st^0)$ correction slightly enhances the threshold $A_\text{th}(\st)$, the rate at which trajectories exceed $A_\text{th}(\st)$ nevertheless increases as $\st$ increases, because $\sigma_S^2 \sim \st^2$ in \Eqnref{transAmat}.
Recently, B\"atge {\em et al.}~\cite{Bat22} described a one-dimensional model for caustic formation where the time needed for caustics to form is accounted for by a $\st$-dependent pre-exponential factor, instead of a $\st$-correction to the action $S(\mbb{A}^*)$ as introduced here.

We make the ansatz
\algn{\eqnlab{Athresh}
	\mbb{A}^*=A_\text{th}(\st)\pmat{1/2&&\\&1/2&\\&&-1}
}
for the gradient matrix, i.e., $A^*_8=\sqrt{3}A_\text{th}(\st)/2$, $A^*_1=\dots=A^*_7=0$, leading to the $\st$-dependent threshold line
\algn{\eqnlab{bifline2}
	 Q = 4R-A_\text{th}^2(\st)\left[A_\text{th}(\st)+3/4\right]\,,
}
which reduces to \Eqnref{bifline} as $\st\to0$. The magnitude of the difference $A_\text{th}(\st)-1/4$ determines by how much the gradient threshold $1/4$ is exceeded at the time $t=t_\text{th}$ when the gradient excursion peaks, and through the dynamics~\eqnref{model}, it also fixes the time $t_\text{c}>t_\text{th}$ at which the caustic is generated. Using this information and input from numerical simulations, we model the optimal fluctuation to next-to-leading order in $\st$.
\subsection{Optimal path to caustic formation}
In order to find the optimal path to caustic formation, we must understand how the optimal fluctuation connects typical values of $\mbb{A}\sim\st$ to the optimal threshold configuration $\mbb{A}^*\sim1$ as a function of time. To make this connection, we use that the most likely way $A^*_i(t)$ for a Gaussian process $A_i(t)$ to reach a large value $A_i\gg\sigma_i$ at a given time $t$ is equal to the correlation function~\eqnref{oscorr} of the process, normalised to the threshold value~\cite{Mei21a}, i.e.,
\algn{\eqnlab{optfluct}
	A^*_i(t) = A_i f_i(t-t_\text{th})\,.
}
We give a derivation of this formula in the Appendix. Since $A^*_8\gg\sigma_S$ for the optimal threshold configuration~\eqnref{Athresh}, we may use \Eqnref{optfluct} to determine the optimal fluctuation $A_i^*(t)$ of the fluid velocity gradients to reach the threshold. This leads to $A^*_8(t) = \sqrt{3}A_\text{th}(\st)f_S(t-t_\text{th})/2$ and $A^*_1(t)=\ldots=A^*_7(t)=0$. Denoting by $\lambda_1(t)>\lambda_2(t)>\lambda_3(t)$ the ordered eigenvalues of $\mbb{A}^*(t)$ along the optimal fluctuation, i.e., 
\algn{
	\mbb{A}^*(t)=\pmat{\lambda_1(t)&&\\&\lambda_2(t)&\\&&\lambda_3(t)}\,,
}
we obtain
\algn{\eqnlab{lamsols}
	\lambda_1(t) = \lambda_2(t) = \frac12 A_\text{th}(\st)f_S(t-t_\text{th})\,,\qquad \lambda_3(t) = -A_\text{th}(\st)f_S(t-t_\text{th})
}
so that $\mbb{A}^*(t_\text{th}) = \mbb{A}^*$ at the time the threshold is reached [c.f. \Eqnref{Athresh}]. The optimal path to caustic formation $\mbb{Z}^*(t)$ for the particle-velocity gradients is now obtained by substituting $\mbb{A}^*(t)$ into \Eqnref{model}. Since $\mbb{A}^*(t)$ is vorticity-free and diagonal in the chosen coordinates, $\mbb{Z}^*(t)$ is initially diagonal, $\mbb{Z}^*(t_0)=\mbb{A}^*(t_0)$. The initial time $t_0$ is chosen such that $|t_\text{c}-t_0|$ is of the order of the expected time for caustic formation, which implies $|t_\text{c}-t_0|\sim\exp[S(\mbb{A}^*)/\st^2]\gg1$ for the model considered here. Since~\Eqnref{model} decouples for diagonal matrices,
\algn{
	\mbb{Z}^*(t) = \pmat{\zeta_1(t)&&\\&\zeta_2(t)&\\&&\zeta_3(t)}
}
remains diagonal along the optimal path to caustic formation. The dynamics of the eigenvalues $\zeta_i(t)$ is then governed by the equations
\algn{\eqnlab{zieqs}
	\st\, \dot \zeta_i(t) = -\zeta_i(t)-\zeta_i^2(t) + \lambda_i(t)\,,
}
with $i=1,2,3$. In words, along the optimal fluctuation, the nine-dimensional matrix equation~\eqnref{model} reduces to three uncoupled equations for the eigenvalues $\zeta_i$, akin to the one-dimensional equation studied in Refs.~\cite{Wil03,Der07}.
The theory summarised here demonstrates that the process of caustic formation in the persistent limit is essentially one dimensional, as shown for two dimensions in Ref.~\cite{Mei21a} and as argued in Ref.~\cite{Bat22}. We also see that the leading negative eigenvalue of $\mbb{A}^*$ must be smaller than $-1/4$ (in our dimensionless formulation) when $\st\to0$.

We solve \Eqsref{zieqs} numerically for a given normalised strain correlation function $f_S(t)$ [see \Eqnref{lamsols}] that we extract from numerical simulations (\Secref{sims}). The threshold value $A_\text{th}(\st)$ is then obtained from \eqnref{zieqs} by tuning $A_\text{th}(\st)$ to match the time difference $t_\text{c}-t_\text{th}(\st)$  for the optimal fluctuation with the numerical value. The optimal fluctuation in the $Q$-$R$ plane reads
\algn{\eqnlab{QReqs}
	Q(t) = -\frac{3}4 A_\text{th}(\st)^2 f_S^2[t-t_\text{th}(\st)]\,,\qquad R(t) = \frac14 A_\text{th}(\st)^3 f_S^3[t-t_\text{th}(\st)]\,.
}
For $\st\to0$, we have $|t_\text{c}-t_0|\to\infty$ and $A_\text{th}(\st)\to1/4$ so that \Eqsref{QReqs} in this case represent an infinite-time trajectory that connects the green and red squares in \Figref{phasediag}, along the right branch of the Vieillefosse line. For finite $\st$, the optimal fluctuation remains on this line, but it now penetrates the yellow region in \Figref{phasediag} for a short time, because $A_\text{th}(\st)>1/4$, and relaxes back to the origin for $t>t_\text{th}$.

Note that the Vieillefosse line separates regions in the $Q$-$R$ plane where $\mbb{A}$ has real eigenvalues, from regions where two eigenvalues form a complex conjugate pair \cite{Cho90}. Thus, on the Vieillefosse line, all eigenvalues are real, and two eigenvalues must be degenerate. Our theory predicts that caustics form when a single eigenvalue of $\mbb{A}$ becomes large and negative, while the other two remain degenerate and positive, with half the magnitude of the negative eigenvalue. For this reason and because the large negative eigenvalue implies $R>0$, this entails that the optimal fluctuation must propagate along the positive branch of the Vieillefosse line for small $\st$.

\section{Numerical simulations}\seclab{sims}
\begin{figure}
	\includegraphics[width=\linewidth]{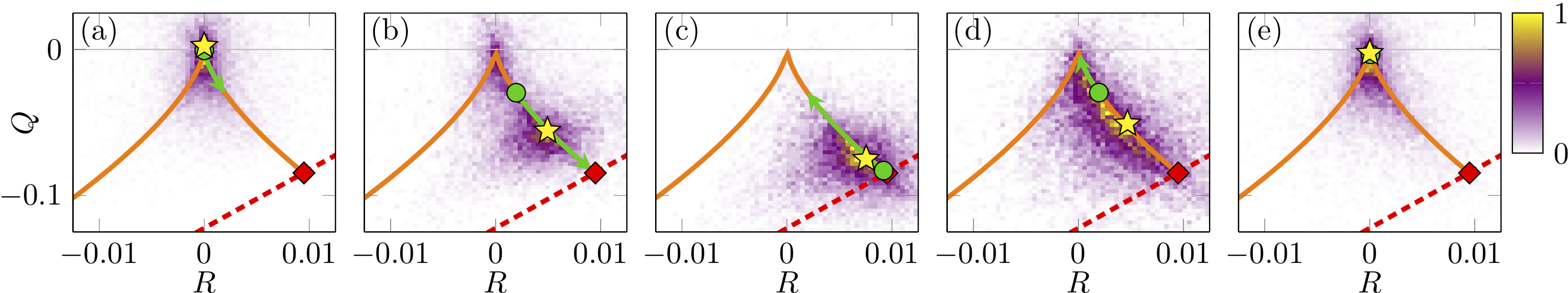}
	\caption{Snapshots of the trajectory density (heat map) in the $Q$-$R$ plane at different times $t-t_\text{c}=-6$ (a), $-4$ (b), $-2.5$ (c), $-1.25$ (d) and $0$ (e) for $\st=0.3$ and $\ku\approx22$. The yellow star shows the location of maximum density. The Vieillefosse line is shown in orange. The red dashed line shows the $\st$-dependent threshold, \Eqnref{bifline2} with $A_\text{th}(\st)\approx0.336$, featuring the optimal threshold configuration (red diamond). The green bullet shows the location of the optimal fluctuation from theory with arrows indicating the change with respect to the next snapshot.}\figlab{phasediag_dens}
\end{figure}
\begin{figure}
	\includegraphics[]{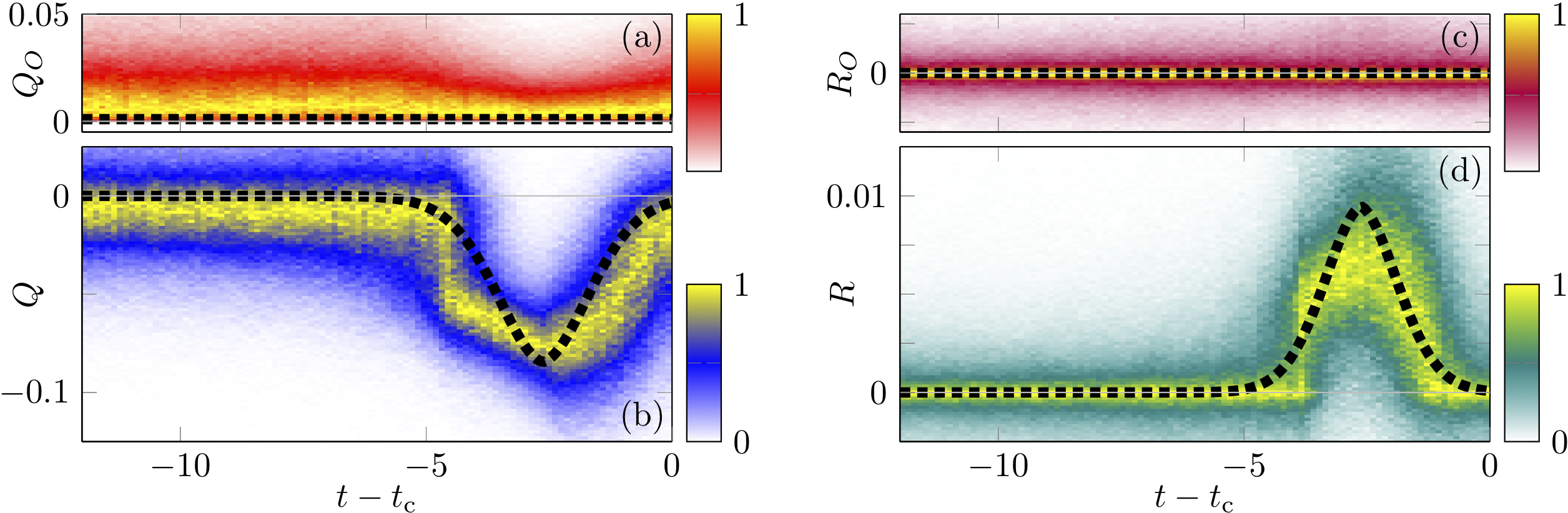}
	\caption{Trajectory densities (heat maps) of (b) $Q$ and (d) $R$ and of their respective vorticity contributions (a) $Q_O$ and (c) $R_O$, normalised to unity at each time slice, as functions of $t$ prior to caustic formation for $\st=0.3$ and $\ku\approx22$. Yellow regions correspond to high trajectory density. The black dotted line shows the optimal fluctuation obtained from theory [\Eqnref{QReqs}].}\figlab{RQ_dens}
\end{figure}
We compare our theory with results of numerical simulations of a Gaussian statistical model for turbulent aerosols at small but finite $\st$. The model captures particle clustering and caustic formation of aerosols in the dissipative range of turbulence qualitatively and, in some cases, even quantitatively~\cite{Gus16}. For the simulations, we evolve a large number of particles with \Eqnref{stokes}. The particles move in a turbulent fluid-velocity field $\ve u(\ve x,t)$, modelled by a fluctuating, statistically homogeneous and isotropic Gaussian vector field with a prescribed correlation function~\cite{Gus16}. In addition, we evolve~\Eqnref{model} for $\mbb{Z}(t)$ along the particle trajectories and track the histories of $\mbb{A}(t)$. Whenever $\tr\mbb{Z}(t)$ reaches a large negative threshold, i.e., a caustic is imminent, we record a caustic and store the corresponding time series of $\mbb{A}$. This provides us with an ensemble of fluid-velocity gradients, conditioned on observing a caustic at each end point. From this ensemble, we evaluate the invariants $Q(t)$ and $R(t)$ and compute numerical approximations of their trajectory densities, allowing us to quantify the most likely fluctuations that result in a caustic at time~\mbox{$t=t_\text{c}$}.

Figure~\figref{phasediag_dens} shows snapshots of the trajectory density in the $Q$-$R$ plane, at different times. Regions of high density are shown in yellow with the maximum marked by a yellow star. The theoretically obtained optimal fluctuation is shown by the green circle. As predicted by the theory, the yellow star performs a large excursion along the positive branch of the Vieillefosse line towards the $\st$-dependent threshold [\Figsref{phasediag_dens}(a) and \figref{phasediag_dens}(b)], reaches it in \Figref{phasediag_dens}(c), and returns to the origin [\Figref{phasediag_dens}(d)]. At the time of caustic formation in \Figref{phasediag_dens}(e), the gradients have almost completely relaxed. Note that typical fluctuations of the fluid-velocity gradients in the $Q$-$R$ plane [\Figref{phasediag_dens}(a)] are tiny compared to the magnitude of the gradient excursion in \Figref{phasediag_dens}(c). Although the theory explains the excursion qualitatively, we observe slight deviations in the time dependences of the green circle and the yellow star.

The time evolutions of the optimal fluctuation and of its vorticity contributions are analysed in more detail in \Figref{RQ_dens}. Vorticity $\mbb{O}$ contributes to $Q$ and $R$ through the expressions $Q_O = \tr(\mbb{O}^{\!\sf T}\!\!\mbb{O})/2$ and $R_O = \tr(\mbb{O}^{\!\sf T}\!\!\mbb{O}\mbb{S})$, respectively. Figures~\figref{RQ_dens}(a) and \figref{RQ_dens}(c) show the vorticity contributions $Q_O$ and $R_O$ as functions of time, together with the theoretical curves for their optimal fluctuations (dotted lines), which are identically zero. The yellow streaks of high trajectory density for $Q_O$ and $R_O$ remain close to zero at all times, which confirms that in both theory and numerics the optimal fluctuation is dominated by the strain $\mbb{S}$ while vorticity $\mbb{O}$ remains small.

Figures~\figref{RQ_dens}(b) and \figref{RQ_dens}(d) show the trajectory densities of $Q$ and $R$ (including both strain and vorticity) as functions of time, together with the theoretical curves for the optimal fluctuations (dotted lines). The yellow streaks for $Q$ and $R$ perform large excursions, centred at $t-t_\text{c}\approx -2.5$, which are in good agreement with our theory. However, the optimal fluctuations in the numerics (yellow streaks) grow faster and are slightly more persistent than predicted. These deviations are likely due to too large $\st$, which leads to inaccuracies in both the optimal fluctuation approach and our approximations.

Outside the regime $\st\ll1$, many fluctuations, not only the optimal one, contribute to caustic formation. Furthermore, the fluid-velocity gradients measured along particle trajectories acquire non-Gaussian statistics and more importantly the time-scale separation in the dynamics of $\mbb{Z}$ and $\mbb{A}$ becomes less sharp. However, the comparison between numerics and theory in \Figsref{phasediag_dens} and \figref{RQ_dens} shows that optimal fluctuation theory does explain caustic formation qualitatively, even at moderate $\st$. The theory not only reproduces the large excursion of the fluid-velocity gradient along the Vieillefosse line, but also correctly predicts vanishing vorticity, as well as the main features in the time dependence of the optimal fluctuation.
\section{Conclusions}
We explained caustic formation in turbulent aerosols for small particle inertia ($\st\ll1$) in three spatial dimensions by means of optimal fluctuation theory. We found that caustics are formed by an instability of spatial particle neighbourhoods that occurs when the fluid-velocity gradients exceed a large threshold in the $Q$-$R$ plane. The most likely way to reach this threshold is by an optimal fluctuation of the fluid-velocity gradients consisting of a large excursion of the strain $\mbb{S}$, while vorticity $\mbb{O}$ remains small. We determined the shape of the excursion explicitly within a mean-field (Gaussian) approximation. The theory predicts that the optimal fluctuation propagates along the positive branch of the Vieillefosse line~\cite{Vie82,Vie84,Can92} to reach the threshold, before it relaxes back to the origin. Using numerical simulations of a Gaussian statistical model, we showed that the optimal fluctuation is dominant not only at infinitesimal $\st$, but also at values of order $\st\sim10^{-1}$, where we observed qualitative agreement with the theory.

The predominance of the strain part in the optimal fluctuation qualitatively explains why collisions are preceded by strong strains and low vorticity in simulations~\cite{Per14}. More generally, our fixed-point analysis shows that at small $\st$, the particle dynamics enters caustic formation only by posing a threshold for the fluid-velocity gradients. This implies that the rate of caustic formation may be estimated solely from the threshold probability~\eqnref{transAmat}, without explicitly referring to the particle dynamics, rather than following local particle neighbourhoods over long times, as in Refs.~\cite{Fal07c,Bha22,Bew13}.

Finally, our analysis indicates that the tear-shaped elongation of the probability distribution of turbulent fluid-velocity gradients along the positive branch of the Vieillefosse line~\cite{Can93,Che99,Che06,Men11b} may facilitate caustic formation in homogeneous and isotropic turbulence, as it leads to an increased probability of reaching the threshold, perhaps even more so if one takes into account preferential concentration~\cite{Joh17}.
\begin{acknowledgments}
K.G. thanks J. Vollmer and G. Bewley for discussions regarding the role of the invariants $Q$ and $R$ for caustic formation.
We thank D. Mitra for discussions regarding caustic formation in turbulence and A. Bhatnagar for sharing his unpublished data~\cite{Bha20} which indicates that also in homogeneous and isotropic turbulence caustics form near the Vieillefosse line at small $\st$, as predicted by our theory. J.M. is funded by a Feodor-Lynen Fellowship of the Alexander von Humboldt-Foundation. B.M. was supported in part by Vetenskapsrådet (VR), Grant No. 2021-4452.  The statistical-model simulations were conducted using the resources of HPC2N provided by the Swedish National Infrastructure for Computing (SNIC), partially funded by VR through Grant No. 2018-05973.
\end{acknowledgments}
\appendix
\section{Derivation of \texorpdfstring{\Eqnref{optfluct}}{Eq. (34)}}\seclab{derivation}
Equation~\eqnref{optfluct} was derived previously in Ref.~\cite{Mei21a}; here we give a simplified derivation. We start with a path-integral expression for \Eqnref{transAi},
\algn{\eqnlab{pathint}
	P[A_i(t_\text{th}) = A_i] = \langle \delta[A_i(t_\text{th})-A_i]\rangle_p = \int\!\!\!\ms{D}A_i(t)\exp\left\{-\frac{\ms{S}[A_i(t)]}{\sigma_i^2}\right\}\delta[A_i(t_\text{th}) - A_i]\,,
}
where $\delta(x)$ denotes the Dirac delta function. For Gaussian processes, the action $\ms{S}[A_i(t)]$ is explicitly known~\cite{Kam07}
\algn{
	\ms{S}[A_i(t)] = \frac12\int_{-\infty}^\infty\!\!\!\ed s \int_{-\infty}^\infty\!\!\!\ed\, t A_i(s) f_i^{-1}(s-t) A_i(t)\,.
}
Here, $f^{-1}(x)$ is the inverse of the normalised correlation function $f_i(x)$ of $A_i(t)$, defined as
\algn{\eqnlab{finverse}
	\int_{-\infty}^\infty\!\!\!\ed s' f_i^{-1}(s-s') f_i(s'-t) = \delta(s-t)\,.
}
For $A_i\gg\sigma_i$, the path integral in \Eqnref{pathint} can be approximated by the saddle point of the action $\ms{S}[A_i(t)]$ under the constraint that $A_i(t_\text{th}) = A_i$ [enforced by the delta function in \Eqnref{pathint}]. We perform the constrained minimisation by adding a Lagrange multiplier $\lambda$ to $\ms{S}[A_i(t)]$:
\algn{\eqnlab{slam}
	\ms{S}_\lambda[A_i(t)] = \frac12\int_{-\infty}^\infty\!\!\!\ed s \int_{-\infty}^\infty\!\!\!\ed\, t A(s) f_i^{-1}(s-t) A(t) - \lambda \left[\int_{-\infty}^\infty\!\!\ed s\, \delta(t_\text{th}-s)A(s) - A_i\right]\,.
}
For computational convenience, we have expressed the constraint $A_i(t_\text{th}) = A_i$ as an integral involving a Dirac delta function, as was done in a different context in Ref.~\cite{Mee22}. We now compute the optimal fluctuation $A^*_i(t)$ as the constrained saddle point $\delta \ms{S}_\lambda[A^*_i(t)]=0$ of the path integral in \Eqnref{pathint}. The variation of \Eqnref{slam} over $A_i(t)$ gives
\algn{
	\delta \ms{S}_\lambda[A^*_i(t)] = \int_{-\infty}^\infty\!\!\!\ed s \,\delta A_i(s) \left[ \int_{-\infty}^\infty\!\!\!\ed t\,f_i^{-1}(s-t) A_i^*(t) - \lambda \delta(t_\text{th}-s)\right]\,,
}
which leads to the saddle point equation
\algn{
	\int_{-\infty}^\infty\!\!\!\ed t \,f_i^{-1}(s-t) A_i^*(t) = \lambda \delta(t_\text{th}-s)\,.
}
Comparison with \Eqnref{finverse} immediately gives
\algn{
	A_i^*(t) = \lambda f_i(t-t_\text{th})\,.
}
The Lagrange parameter $\lambda$ is now simply obtained by using the constraint $A_i^*(t_\text{th})=A_i$ and the normalisation condition $f_i(0)=1$. This gives $\lambda = A_i$ and thus the optimal fluctuation~\eqnref{optfluct}. We evaluate the action $\ms{S}[A_i(t)]$ at the saddle point by using the optimal fluctuation~\eqnref{optfluct} and obtain
\algn{
	\ms{S}[A^*_i(t)] = \frac{A_i^2}2\,.
}
Hence, we recover the quadratic action in \Eqnref{transAi} from \Eqnref{pathint}, evaluated at the optimal fluctuation~\eqnref{optfluct}.
\end{document}